\documentclass[aps,prl,twocolumn,superscriptaddress]{revtex4-2}

\usepackage{graphicx}
\usepackage{wrapfig}
\usepackage{soul}
\usepackage{xcolor}
\begin{document}



\title{First CLAS12 measurement of DVCS beam-spin asymmetries in the extended valence region}







\newcommand*{\SACLAY}{IRFU, CEA, Universit\'{e} Paris-Saclay, F-91191 Gif-sur-Yvette, France}
\newcommand*{\GLASGOW}{University of Glasgow, Glasgow G12 8QQ, United Kingdom}
\newcommand*{\ANL}{Argonne National Laboratory, Argonne, Illinois 60439}
\newcommand*{\CSUDH}{California State University, Dominguez Hills, Carson, CA 90747}
\newcommand*{\CANISIUS}{Canisius College, Buffalo, NY}
\newcommand*{\CUA}{Catholic University of America, Washington, D.C. 20064}
\newcommand*{\CNU}{Christopher Newport University, Newport News, Virginia 23606}
\newcommand*{\UCONN}{University of Connecticut, Storrs, Connecticut 06269}
\newcommand*{\DUKE}{Duke University, Durham, North Carolina 27708-0305}
\newcommand*{\DUQUESNE}{Duquesne University, 600 Forbes Avenue, Pittsburgh, PA 15282 }
\newcommand*{\FU}{Fairfield University, Fairfield CT 06824}
\newcommand*{\FERRARAU}{Universita' di Ferrara , 44121 Ferrara, Italy}
\newcommand*{\FIU}{Florida International University, Miami, Florida 33199}
\newcommand*{\GWUI}{The George Washington University, Washington, DC 20052}
\newcommand*{\GSIFFN}{GSI Helmholtzzentrum fur Schwerionenforschung GmbH, D 64291 Darmstadt, Germany}
\newcommand*{\INFNFE}{INFN, Sezione di Ferrara, 44100 Ferrara, Italy}
\newcommand*{\INFNFR}{INFN, Laboratori Nazionali di Frascati, 00044 Frascati, Italy}
\newcommand*{\INFNGE}{INFN, Sezione di Genova, 16146 Genova, Italy}
\newcommand*{\INFNRO}{INFN, Sezione di Roma Tor Vergata, 00133 Rome, Italy}
\newcommand*{\INFNTUR}{INFN, Sezione di Torino, 10125 Torino, Italy}
\newcommand*{\INFNPAV}{INFN, Sezione di Pavia, 27100 Pavia, Italy}
\newcommand*{\ORSAY}{Universit\'{e} Paris-Saclay, CNRS/IN2P3, IJCLab, 91405 Orsay, France}
\newcommand*{\Juelich}{Institute fur Kernphysik (Juelich), Juelich, Germany}
\newcommand*{\JMU}{James Madison University, Harrisonburg, Virginia 22807}
\newcommand*{\KNU}{Kyungpook National University, Daegu 41566, Republic of Korea}
\newcommand*{\LAMAR}{Lamar University, 4400 MLK Blvd, PO Box 10046, Beaumont, Texas 77710}
\newcommand*{\MIT}{Massachusetts Institute of Technology, Cambridge, Massachusetts  02139-4307}
\newcommand*{\MISS}{Mississippi State University, Mississippi State, MS 39762-5167}
\newcommand*{\ITEP}{National Research Centre Kurchatov Institute - ITEP, Moscow, 117259, Russia}
\newcommand*{\UNH}{University of New Hampshire, Durham, New Hampshire 03824-3568}
\newcommand*{\NMSU}{New Mexico State University, PO Box 30001, Las Cruces, NM 88003, USA}
\newcommand*{\NSU}{Norfolk State University, Norfolk, Virginia 23504}

\newcommand*{\OHIOU}{Ohio University, Athens, Ohio  45701}

\newcommand*{\ODU}{Old Dominion University, Norfolk, Virginia 23529}

\newcommand*{\JLUGiessen}{II Physikalisches Institut der Universitaet Giessen, 35392 Giessen, Germany}

\newcommand*{\URICH}{University of Richmond, Richmond, Virginia 23173}

\newcommand*{\ROMAII}{Universita' di Roma Tor Vergata, 00133 Rome Italy}

\newcommand*{\MSU}{Skobeltsyn Institute of Nuclear Physics, Lomonosov Moscow State University, 119234 Moscow, Russia}

\newcommand*{\SCAROLINA}{University of South Carolina, Columbia, South Carolina 29208}
\newcommand*{\TEMPLE}{Temple University,  Philadelphia, PA 19122 }
\newcommand*{\JLAB}{Thomas Jefferson National Accelerator Facility, Newport News, Virginia 23606}
\newcommand*{\UTFSM}{Universidad T\'{e}cnica Federico Santa Mar\'{i}a, Casilla 110-V Valpara\'{i}so, Chile}
\newcommand*{\INSUBRIA}{Universit\`{a} degli Studi dell'Insubria, 22100 Como, Italy}
\newcommand*{\BRESCIA}{Universit`{a} degli Studi di Brescia, 25123 Brescia, Italy}
\newcommand*{\UCR}{University of California Riverside, 900 University Avenue, Riverside, CA 92521, USA}
\newcommand*{\YORK}{University of York, York YO10 5DD, United Kingdom}
\newcommand*{\VIRGINIA}{University of Virginia, Charlottesville, Virginia 22901}
\newcommand*{\WM}{College of William and Mary, Williamsburg, Virginia 23187-8795}
\newcommand*{\YEREVAN}{Yerevan Physics Institute, 375036 Yerevan, Armenia}

\newcommand*{\NOWANL}{Argonne National Laboratory, Argonne, Illinois 60439}

\author{G.~Christiaens}
\affiliation{\SACLAY}
\affiliation{\GLASGOW}
\author{M.~Defurne}
\altaffiliation[contact author : ]{maxime.defurne@cea.fr}
\affiliation{\SACLAY}
\author {D.~Sokhan} 
\affiliation{\SACLAY}
\affiliation{\GLASGOW}
\author {P.~Achenbach} 
\affiliation{\JLAB}
\author {Z.~Akbar} 
\affiliation{\VIRGINIA}
\author {M.J.~Amaryan} 
\affiliation{\ODU}
\author {H.~Atac} 
\affiliation{\TEMPLE}
\author {H.~Avakian}
\affiliation{\JLAB}
\author {C. Ayerbe Gayoso} 
\affiliation{\WM}
\author {L.~Baashen} 
\affiliation{\FIU}
% ...
\collaboration{CLAS collaboration}

\date{\today}

\begin{abstract}
Deeply virtual Compton scattering (DVCS) allows one to probe Generalized Parton Distributions (GPDs) describing the 3D structure of the nucleon. We report the first measurement of the DVCS beam-spin asymmetry using the CLAS12 spectrometer with a 10.2 and 10.6 GeV electron beam scattering from unpolarised protons. The results greatly extend the $Q^2$ and Bjorken-$x$ phase space beyond the existing data in the valence region and provide over 2000 new data points measured with unprecedented statistical uncertainty, setting new, tight constraints for future phenomenological studies. 
\end{abstract}         


\maketitle

\section{Introduction}

Generalized Parton Distributions (GPDs) serve as a powerful tool for describing the three-dimensional dynamics of nucleon structure, including the composition of spin and pressure distributions \cite{Muller94, Raduyshkin96, Ji97, Burkardt00, Polyakov03}. 
These functions can be accessed via electron scattering, where the squared four-momentum transfer to a parton, $Q^2$, results in a change -2$\xi$ in its longitudinal momentum fraction $x$, and a change in the nucleon's momentum after reabsorption of the parton, contained in the Mandelstam variable $t$. 

The amplitudes of deeply virtual Compton scattering
\begin{wrapfigure}{r}{0.25\textwidth}
  \vspace{-20pt}
  \begin{center}
    \includegraphics[width=0.25\textwidth]{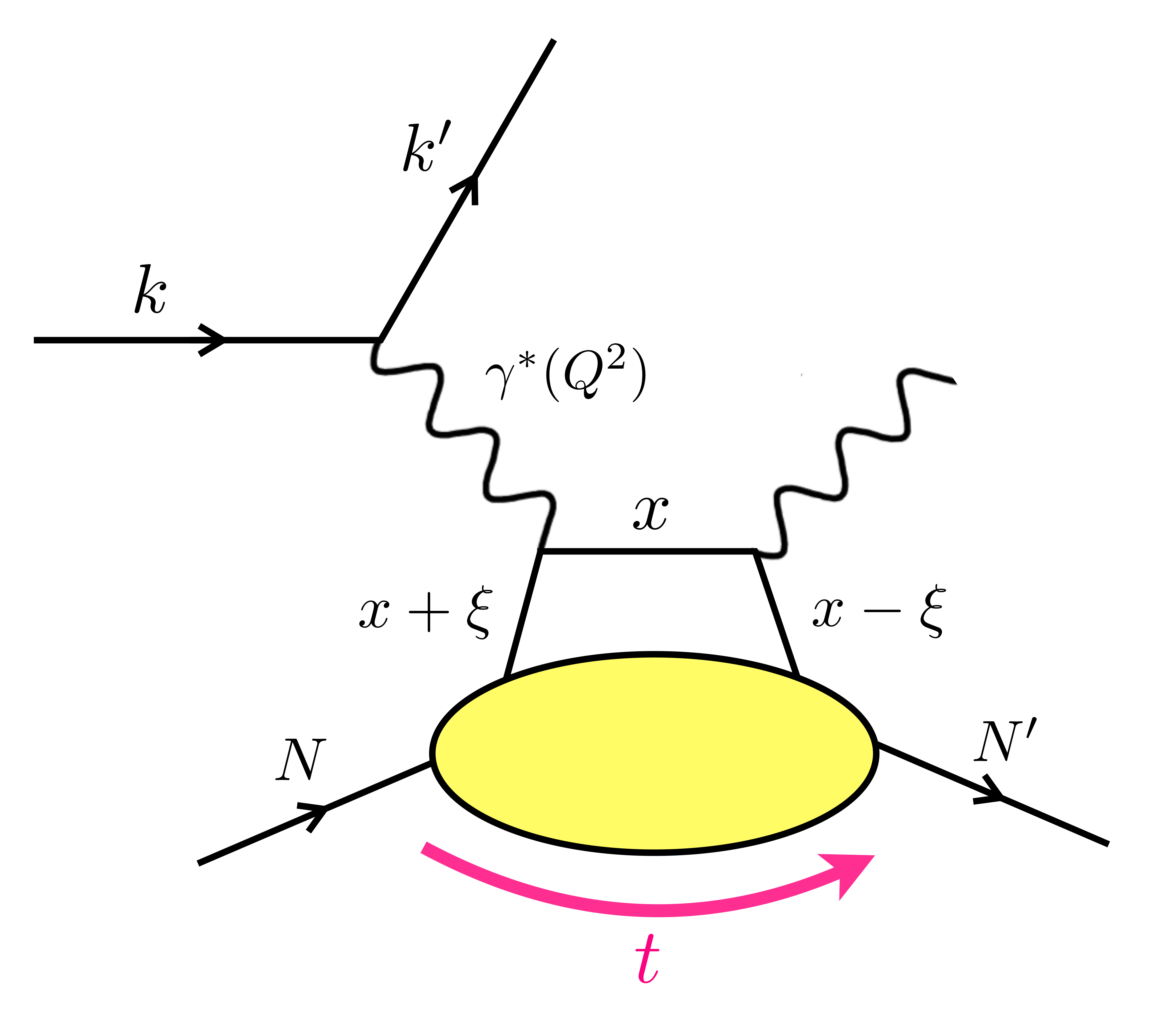}
  \end{center}
  \vspace{-20pt}
  \caption{The DVCS process.}
  \vspace{-10pt}
  \label{fig:fig1a}
\end{wrapfigure}
 (DVCS), characterized by the struck parton emitting a high-energy photon (Fig.\ref{fig:fig1a}) and the nucleon remaining intact, are parametrized by Compton Form Factors (CFFs), complex functions that are $x$-integrals of the corresponding GPDs.
 The process is sensitive, at high $Q^2$, low $t$ and leading order in perturbative Quantum Chromodynamics (pQCD), to the CFFs corresponding to four GPDs: $H$, $E$, $\tilde{H}$ and $\tilde{E}$ \cite{Muller94, Raduyshkin96, Ji97, Burkardt00}. In the experimentally indistinguishable process, Bethe-Heitler (BH), the photon is instead radiated by the incoming or scattered lepton. As BH and DVCS share the same final state, the observed photon electroproduction results from the interference of both processes and the cross section can be written as:
\begin{equation}
\sigma^{e^\pm p\gamma}=\sigma^{BH}+\sigma^{DVCS}\mp \left( \mathcal{I}^{Re} + \lambda_e \mathcal{I}^{Im}  \right),
\end{equation}
where $\lambda_e$ is the beam polarization, $\sigma^{BH}$ and $\sigma^{DVCS}$ are the cross sections of BH and DVCS, respectively, and $\mathcal{I}^{Re}$ and $\mathcal{I}^{Im}$ are the real and imaginary parts of the interference term between the amplitudes for both processes.


The beam-spin asymmetry in photon electroproduction is an experimentally attractive observable, since, to first approximation, detector acceptance effects cancel. Produced mostly by the interference of BH and DVCS, for a proton target in the valence region, it can reach up to $\sim 20\%$ (as observed in the previously collected data) and is dominated by $\text{Im}\mathcal{H}$, the imaginary part of the CFF corresponding to the GPD $H$~\cite{BMK2002,Kroll:2012sm}. 

Exclusive photon electroproduction on the nucleon has been successfully studied at a number of facilities around the world, such as DESY with H1 \cite{H101, H108}, ZEUS \cite{ZEUS03} and HERMES \cite{HERMES}, COMPASS \cite{COMPASS:2018pup,COMPASS} and Jefferson Lab Halls A \cite{JeffersonLabHallA:2006prd, HallA15, HallA17, HallA07, HallA21, HallA_prl} and B \cite{CLAS_DVCS01, CLAS_DVCS06, CLAS_DVCS08, CLAS_DVCS15, CLAS_DVCS15_2, CLAS_DVCS15_3, CLAS_DVCS18}. 
In this article, we report the first DVCS measurements performed with a beam energy of just over 10~GeV and the CLAS12 spectrometer in Hall B at Jefferson Lab. Approximately 85\% of the new data cover a phase space in the valence quark region that has never been probed with DVCS before. The data were collected in Fall 2018 with a beam energy of 10.6~GeV and Spring 2019 at 10.2~GeV. For both run periods, which used an unpolarized liquid-H$_2$ target, the longitudinal beam polarization was $\sim$86\%. This, along with very large statistics, enabled measurements of the beam-spin asymmetry in DVCS off the proton to be made in 64 bins of $x_B$, $Q^2$ and $t$. 


\section{The CLAS12 spectrometer}
The CLAS12 spectrometer \cite{CLAS_NIM} can be decomposed into a central and a forward part. The central part, around the target, is placed in a 5T solenoidal magnetic field, detecting particles emitted at polar angles between 35$^{\circ}$ and 125$^\circ$ with respect to the beam direction. Silicon and Micromegas trackers are used to reconstruct charged tracks while a time-of-flight scintillator detector enables particle identification. At the heart of the forward part are drift chambers placed in a toroidal magnetic field for charged track reconstruction. Electron identification is provided by a high-threshold \v{C}erenkov detector 
complemented with an electromagnetic sampling calorimeter.
Hadrons are identified using a scintillator time-of-flight detector placed between the drift chambers and the calorimeter. 

Scattered electrons are detected in the forward part of CLAS12. About 80\% of the recoil protons are detected in the central part while the remaining 20\% go forward. Finally, the photon is detected either in the electromagnetic calorimeter or in a specialized forward tagger, designed to cover polar angles of 2$^{\circ}$ - 5$^{\circ}$ from the beam direction. 

\section{Beam-spin asymmetry in $ep \rightarrow e'p'\gamma$}
\begin{figure}
    \centering
    \begin{tabular}{c}
         \includegraphics[width=\linewidth]{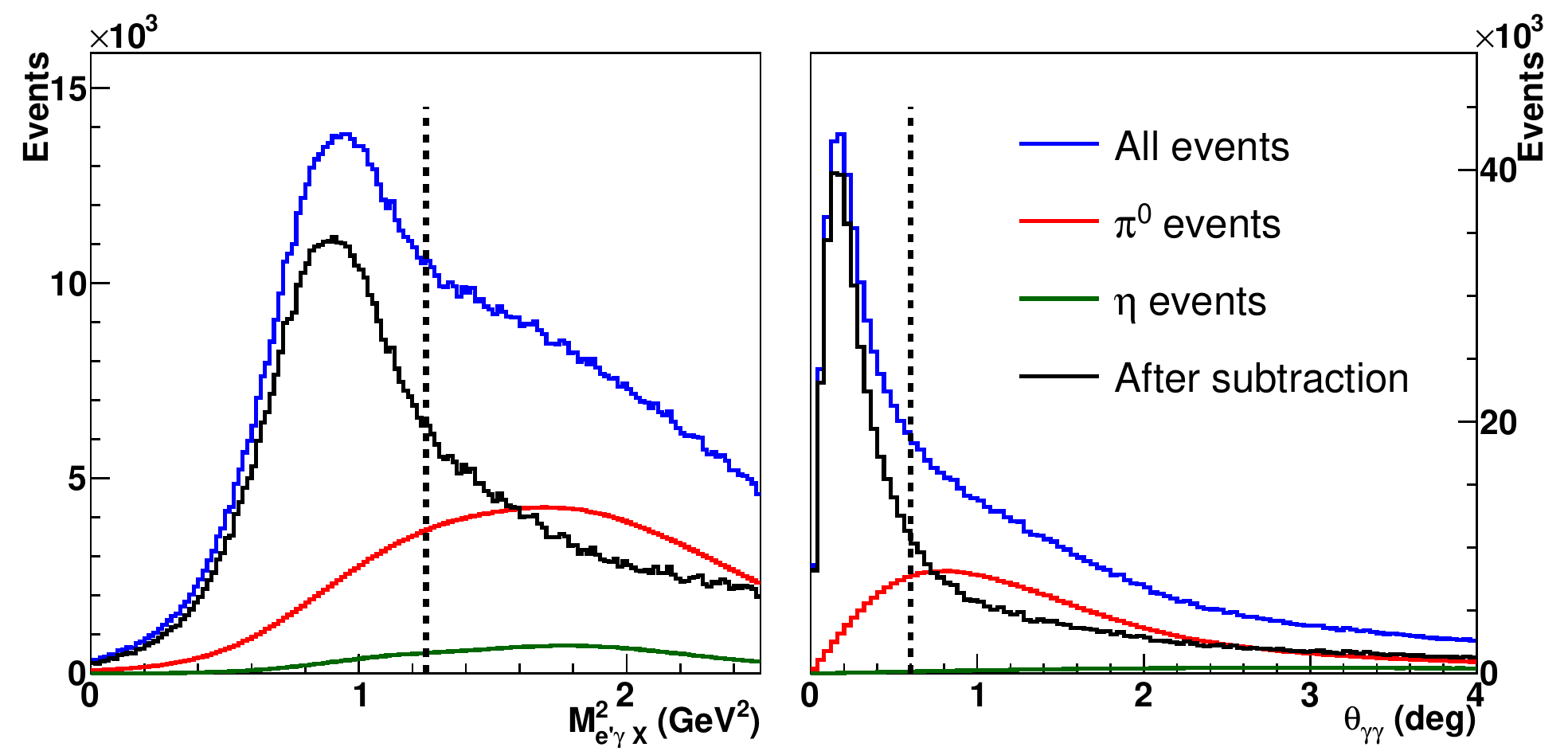}
    \end{tabular}
    \caption{Squared missing mass $M^2_{e'\gamma X}$ (left) and cone angle $\theta_{\gamma\gamma}$ (right) distributions prior to the application of any exclusivity cuts, for the full data-set showing the $\pi^0$ and $\eta$ contamination and after subtraction. Dashed lines indicate the position of the cuts.}
    \label{fig:exclusivity}
\end{figure}

Events with a single high-energy electron, a single proton and at least one photon above 2~GeV were considered as BH / DVCS candidates. The highest-energy photon is selected, if more than one meets the criteria in the event. Exclusivity was ensured by application of cuts on the following variables:
\begin{itemize}
    \item $\theta_{\gamma\gamma}$: the cone angle between the detected photon and the expected photon direction in $ep \rightarrow e'p'\gamma$, derived kinematically using the scattered electron, the detected proton and momentum conservation. It is cut at 0.6$^\circ$ and shown in Fig.~\ref{fig:exclusivity} right.
    \item $E_{miss}$: the ``missing" energy balance between the initial state, $ep$, and the $e'p'\gamma$ final state. It is cut at 0.5~GeV.
    \item $p_{Tmiss}$: the ``missing" transverse (with respect to the beam) momentum balance between the $ep$ and the $e'p'\gamma$ states is cut at 0.125~MeV.
    \item $M^2_{e'\gamma X}$ : the squared ``missing" mass of $X$ in the process $ep \rightarrow e'\gamma X$, which should correspond to a proton for exclusive reconstruction of DVCS or BH and is consequently cut at 1.25~GeV$^2$. The distribution is displayed in Fig.~\ref{fig:exclusivity} left. 
\end{itemize}



The main background to the DVCS-BH process, which is reduced but not entirely eliminated by the exclusivity cuts, comes from $\pi^0$ electroproduction. In this process, instead of a photon, the target proton emits a neutral pion with energy similar to that of a DVCS photon, since the pion mass is low. The exclusive $\pi^0$-production is also sensitive to GPDs and carries its own beam-spin asymmetry, although it is typically on the few-percent level~\cite{Masi_2008} and much lower than that of DVCS-BH.  

In its center-of-mass frame, the pion decays into two back-to-back photons but in the lab frame, their direction and energy depend on the relative orientation of their momenta with respect to the boost direction defined by the pion momentum. When the decay is collinear to the pion momentum, one photon carries almost all of the pion energy and may mimic a DVCS photon, passing the exclusivity cuts.
To subtract this contamination, we applied the technique developed in~\cite{HallA15}. Applying identical selection criteria for the electron and the proton as in the DVCS analysis, a $\pi^0$-sample was created from the data by cutting on the two-photon invariant mass and loosely cutting on $M^2_{e'\gamma \gamma X}$, the squared missing mass associated with $X$ in $ep \rightarrow e'\gamma\gamma X$. This selects a sample of $ep \rightarrow e'p'\pi^0$ events. Next, for each $\pi^0$ in this sample, the associated DVCS contamination was derived by generating 1500 decays of $\pi^0 \rightarrow \gamma \gamma$ and normalizing those of the decays that lead to DVCS contamination, by the number of decays leading to a detected $\pi^0$. The number of decays was optimised between desired statistics and processing time. In the simulation, pair conversion for the photons and calorimeter reconstruction efficiency were both taken into account. Fig. \ref{fig:exclusivity} shows the integrated distributions of two exclusivity variables, the squared missing mass and the cone angle, for the determined $\pi^0$ contamination, the negligible contribution of $\eta$-production (estimated with the same method) and the DVCS-BH sample after meson-background subtraction. 

The $\pi^0$ contamination was determined for each helicity state, thus entangling the $\pi^0$ beam-spin asymmetry in the subtraction from the distributions passing the DVCS selection, which was done on a bin-by-bin basis. 
In addition to the statistical uncertainty induced by the subtraction of the $\pi^0$ events, the systematic uncertainty related to it is given for each bin by:
\begin{equation}
    \Delta A = \frac{\sigma_f \times \left(A^{raw}-A^{\pi}\right)}{\left(1-f\right)^2}\;,
\end{equation}
with $f$ the fraction of contamination, $\sigma_f$ the associated uncertainty, $A^{raw}$ and $A^{\pi}$ the asymmetries prior to $\pi^0$-contamination subtraction and of the subtracted $\pi^0$ contamination, respectively. Using Monte-Carlo simulations with two different $\pi^0$ event generators, we estimated $\sigma_f=0.1\times f$. As the $\pi^0$-statistics may not be high enough to derive its beam-spin asymmetry with accuracy, $A^{\pi}$ was set to 0 for a conservative estimate of the systematics. Since the fraction of contamination depends on exclusivity cuts as well as on the ratio between the DVCS and $\pi^0$ cross sections, and thus varies from bin to bin, the systematic uncertainty was added quadratically to the statistical uncertainty of the DVCS beam-spin asymmetry. 

\begin{wrapfigure}{r}{0.25\textwidth}
  \vspace{-20pt}
  \begin{center}
    \includegraphics[width=0.25\textwidth]{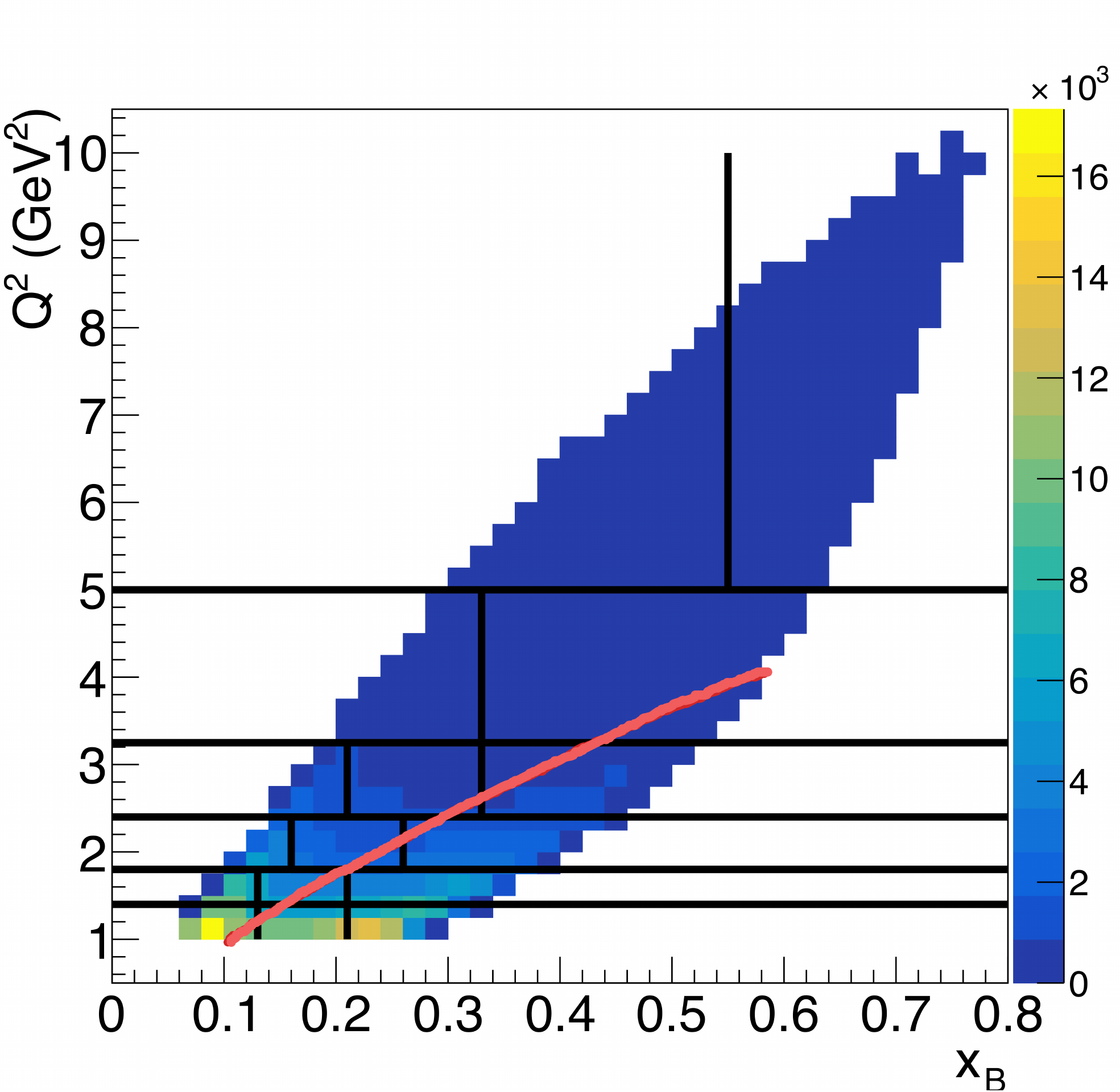}
  \end{center}
  \vspace{-20pt}
  \caption{CLAS12 phase space in $Q^2$ vs. $x_B$, showing the division into 16 bins. The red line indicates the approximate upper reach of CLAS data at 6~GeV. }
  \vspace{-10pt}
  \label{fig:figfeather}
\end{wrapfigure}

The detection of the scattered lepton in $ep \rightarrow e'p'\gamma$ allows one to describe the reaction kinematics in terms of the variables $Q^2=-q^2=-(k-k')^2$ and $x_B=Q^2/(2q\cdot N)$ (see Fig.~\ref{fig:fig1a}). The variables $t$ and $\phi$ (the angle between the leptonic and hadronic planes in the process) were computed using the scattered lepton kinematics and the direction of the photon, the latter being a well-reconstructed quantity.
 As shown in Fig.~\ref{fig:figfeather}, there are 16 bins covering the $Q^2$/$x_B$ phase space. Each $Q^2$/$x_B$-bin was further subdivided into 4 bins in $t_{min}(Q^2,x_B)-t$, with $t_{min}(Q^2,x_B)$ the minimal squared momentum transfer.
An adaptive binning, dictated by the statistics, was implemented for the variable $\phi$, as the cross section exhibits a steep dependence on it. For each $Q^2$/$x_B$/$t$/$\phi$-bin, the averaged kinematic values were computed, corrected for the $\pi^0$ contamination bias. Detector acceptance and fiducial cuts applied for particle selection result in a non-negligible variation of $x_B$ and $t$, which can be observed as a function of $\phi$. Thus, the average kinematics are not necessarily the same for neighbouring $\phi$-bins.  

Radiative effects, where either a soft photon was radiated by the incoming or outgoing electron, or there was a QED loop involving the virtual photon vertex, were considered and corrected for. These effects may result in bin migration -- where the reconstructed kinematics of an event differ from the true kinematics at the vertex.
The bin migration was corrected by deriving a migration matrix from a Monte-Carlo simulation using a DVCS event generator that included soft photon radiation, based on the calculations by Akusevitch \emph{et al.}\cite{Akusevitch}. 

\section{Results}
In total, the beam-spin asymmetry was obtained for 64 bins in $Q^2$, $x_B$ and $t$, each of which contained between 10 and 33 bins in $\phi$, separately for the two data-sets obtained with a beam energy of 10.2 and 10.6~GeV. These datasets provide a tremendous addition to the world data. 
In order to put the results of the CLAS12 datasets into perspective, we will refer to two such studies performing global fits, prior to the inclusion of the current data, using very different approaches. The first study has been performed by Kumericki \emph{et al.} and is based on a GPD-hybrid model~\cite{KM15}, with sea partons described by a Mellin-Barnes partial-wave expansion, while dispersion relation techniques are applied to the valence region. The few parameters of the model are then fitted against most of the DVCS data available, yielding the KM15 model. The second approach, developed by the PARTONS Collaboration, is based on artificial neural networks (ANN)~\cite{PARTONS, PARTONSann}, trained on the world data-set of DVCS measurements. In both the KM15 and ANN methods, the fit in the valence region occurs at the level of Compton Form Factors. As is the case for any neural-network based approach, the ANN method leads to a minimally biased description of the DVCS measurements. To propagate uncertainties, the PARTONS Collaboration smeared the datasets a hundred times according to the quoted systematic uncertainties by the experimental collaborations, thus yielding a library of 100 ANNs whose mean is the fitted CFF value and whose standard deviation provides the CFF uncertainty.

\begin{figure}
    \centering
    \includegraphics[width=\linewidth]{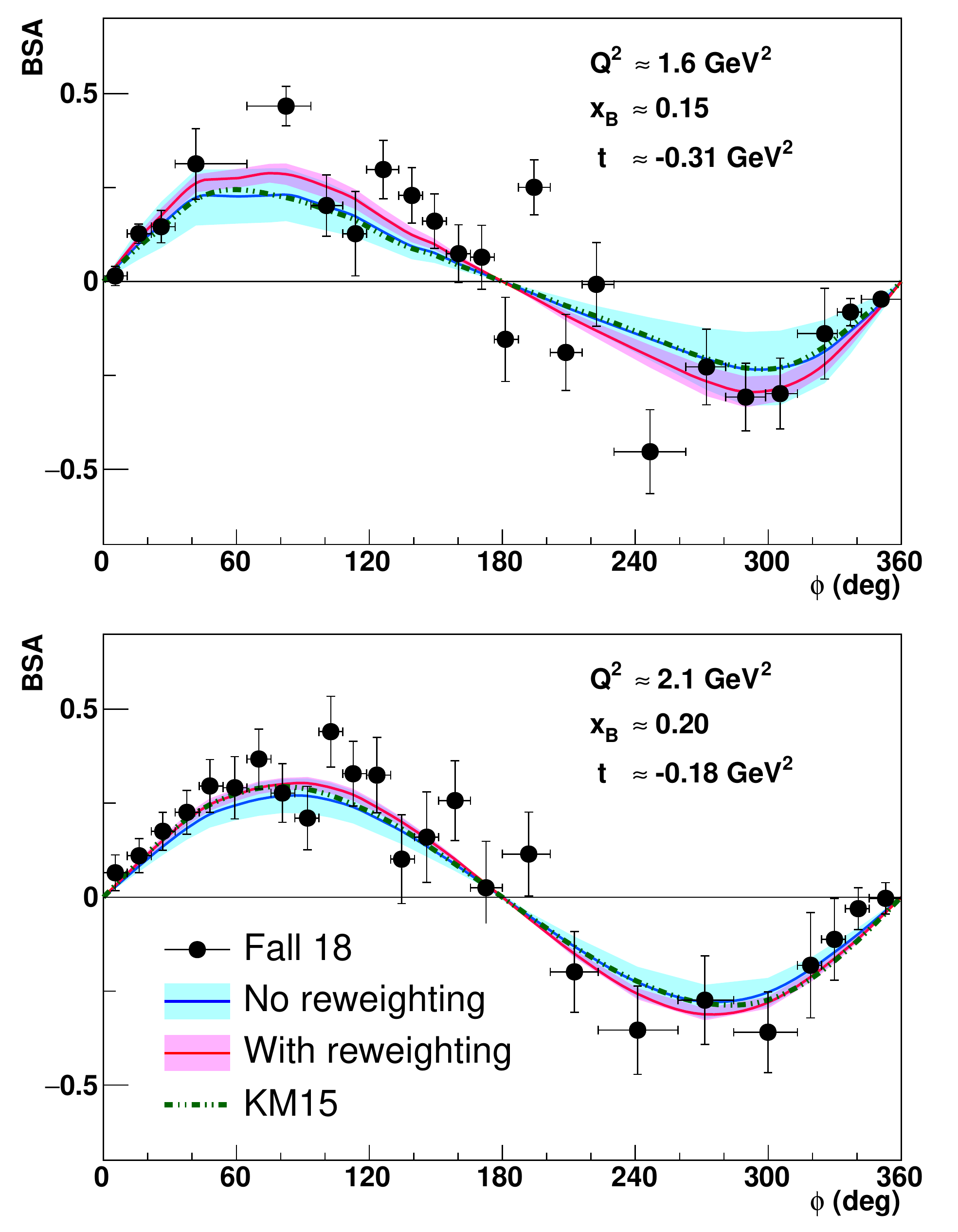}
    \caption{Beam-spin asymmetries in two kinematic bins, compared with PARTONS ANNs before and after reweighting, as well as with KM15. The kinematics listed are approximate -- point-by-point kinematics are available in the tables in the supplemental material.}
    \label{fig:reweighting}
\end{figure}

For bins in phase space that overlap with the ANN training measurements, it is possible to use a Bayesian reweighting technique to test each ANN against the DVCS asymmetries presented in this work. Regularly applied in the PDF field~\cite{Giele_1998}, the technique consists of computing a weight associated with each ANN, which reflects how closely it agrees with the new data. The weight $\omega_k$ associated with the $k$-replica is given by\cite{Dutrieux:2021ehx}: \begin{equation}
\omega_k=\frac{1}{Z}\chi_k^{n-1}e^{-(\chi^2_k/2)}\;,    
\end{equation}
where $Z=\sum_{k} \chi_k^{n-1}e^{-(\chi^2_k/2)}$, $n$ is the number of points used to compute $\chi^2_k=\sum_{n}(y-y_n)V^{-1}(y-y_n)^T$, $y$ are the asymmetry values from CLAS12 with $V$ the associated covariance matrix, and $y_n$ are the predictions of the $k$-replica. By computing the weighted average and standard deviation of the 100 ANNs, the impact of the CLAS12 data can thus be visualized. As outputs of a minimally biased approach, the ANN fits provide a firm constraint on CFFs only within the training phase-space. Therefore, to perform the reweighting, two $Q^2$/$x_B$/$t$-bins that had similar kinematics to previously published CLAS data were chosen. In Fig.~\ref{fig:reweighting}, it can be seen that both KM15 and ANN predictions agree very well but seem to slightly underestimate the asymmetry at $x_B$=0.15. Although in similar kinematics as previous CLAS data, the beam energy of the present dataset is significantly different and may thus bring new constraints to separate the imaginary and real parts of CFFs through a Rosenbluth separation. 

The quantity $N_{eff}$ allows one to estimate an effective number of ANNs contributing significantly to the reweighted average and standard deviation:
\begin{equation}
     N_{eff}=\exp\left(-\sum_{k=1}^{N_{rep}} \omega_k \ln{\omega_k}\right)\;.
\end{equation}
 $N_{eff}$ is found to be $\sim$ 30 in Fig.~\ref{fig:reweighting}, meaning that the precision of the new data points provided by CLAS12 rejects 70\% of the ANNs that are in agreement with the JLab 6-GeV data within the statistical accuracy. In fact, its constraining and highly discriminative power illuminates the necessity for a full global fit of all world data sensitive to CFFs, with the inclusion of this vast, newly collected dataset.

\begin{figure*}
    \centering
    \includegraphics[width=\linewidth]{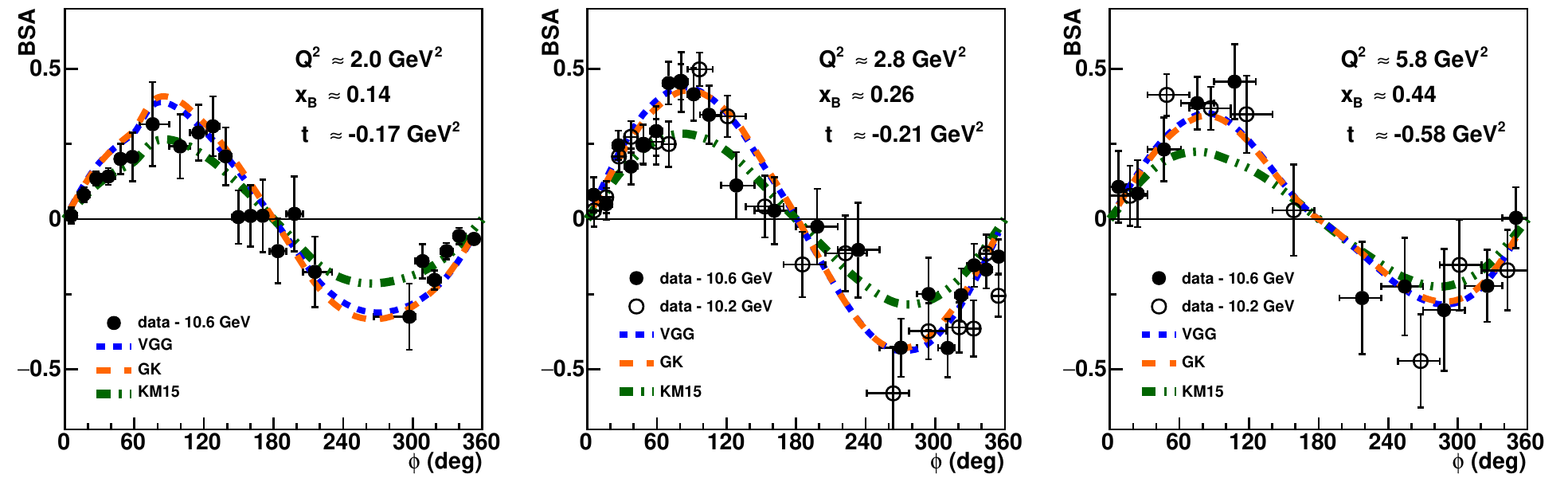}
    \caption{Beam-spin asymmetries for bins only reachable with a $\sim$ 10 GeV electron beam, compared with the KM15, GK and VGG GPD-models. The kinematics listed are approximate -- point-by-point kinematics are available in the tables in the supplementary material for the full data-set.}
    \label{fig:incognita}
\end{figure*}

Figure~\ref{fig:incognita} displays three additional bins in regions of phase space that could only be reached with a 10~GeV beam. Established GPD models such as Goloskokov-Kroll (GK)~\cite{GK05, GK10} and Vanderhaeghen-Guichon-Guidal (VGG)~\cite{VGG_recent,VGG_recent2} describe the new data in the unexplored phase space reasonably well, while KM15 seems to underestimate the amplitude of the asymmetry for some of the new bins.  

\section{Conclusion}
In conclusion, we report the first measurements of the beam-spin asymmetry in deeply virtual Compton scattering with a lepton beam energy of just over 10~GeV. This has extended the explored $Q^2$ and $x_B$ phase space greatly beyond the valence-region measurements done at Jefferson Lab with a 6~GeV beam and will allow results from both beam energies to be studied in a Rosenbluth-like separation to access the purely-DVCS terms in the scattering amplitude.

Although representing only 25\% of the beam-time allocated to the CLAS12 experiment for DVCS on an unpolarized proton, these new results are already statistically competitive with the entire 6-GeV program, as demonstrated by the reweighting technique. In the \emph{terra incognita}, which forms the great majority of the phase-space covered by the new measurement, while GPD models seem to be in fair agreement with the newly collected data, some tension can be seen with the KM15 global fit, illuminating the need for the inclusion of the new data-points. The new data have greatly enriched the world set, extending the probed phase-space in the valence region with high-precision measurements, and promise to provide both very significant constraints for global fits across new kinematic ranges and a crucial means of validating and refining GPD models. 
All data points can be found in supplemental materials associated to this Letter. 


\section{Acknowledgements}
We acknowledge the outstanding efforts of the staff of the Accelerator, the Nuclear Physics Division, Hall B, and the Detector Support Group at JLab that have contributed to the design, construction, installation, and operation of the CLAS12 detector. We thank Maurizio Ungaro for his contributions in the CLAS12 simulations. We also thank the CLAS Collaboration for staffing shifts and taking high quality data. This work was supported by the United States Department of Energy under JSA/DOE Contract DE-AC05-06OR23177. This work was also supported in part by the U.S. National Science Foundation, the State Committee of Science of the Republic of Armenia, the Chilean Agencia Nacional de Investigación y Desarrollo, the Italian Istituto Nazionale di Fisica Nucleare, the French Centre National de la Recherche Scientifique, the French Commissariat a l’Energie Atomique, the Scottish Universities Physics Alliance (SUPA), the United Kingdom Science and Technology Facilities Council (STFC), the National Research Foundation of Korea, the Deutsche Forschungsgemeinschaft (DFG). This research was funded in part by the French Agence Nationale de la Recherche contract no. 37. This work was supported as well by the EU Horizon 2020 Research and Innovation Program under the Marie Sklodowska-Curie Grant Agreement No. 101003460 and the Ile-de-France region via the Blaise Pascal Chair of International Excellence. This work has been funded by the U.K. Science and Technology Facilities Council under grants ST/P004458/1 and ST/V00106X/1 and in part by the Chilean National Agency of Research and Development ANID PIA/APOYO AFB180002.

\bibliography{PRL_bib.bib}

\begin{thebibliography}{38}%
\makeatletter
\providecommand \@ifxundefined [1]{%
 \@ifx{#1\undefined}
}%
\providecommand \@ifnum [1]{%
 \ifnum #1\expandafter \@firstoftwo
 \else \expandafter \@secondoftwo
 \fi
}%
\providecommand \@ifx [1]{%
 \ifx #1\expandafter \@firstoftwo
 \else \expandafter \@secondoftwo
 \fi
}%
\providecommand \natexlab [1]{#1}%
\providecommand \enquote  [1]{``#1''}%
\providecommand \bibnamefont  [1]{#1}%
\providecommand \bibfnamefont [1]{#1}%
\providecommand \citenamefont [1]{#1}%
\providecommand \href@noop [0]{\@secondoftwo}%
\providecommand \href [0]{\begingroup \@sanitize@url \@href}%
\providecommand \@href[1]{\@@startlink{#1}\@@href}%
\providecommand \@@href[1]{\endgroup#1\@@endlink}%
\providecommand \@sanitize@url [0]{\catcode `\\12\catcode `\$12\catcode
  `\&12\catcode `\#12\catcode `\^12\catcode `\_12\catcode `\%12\relax}%
\providecommand \@@startlink[1]{}%
\providecommand \@@endlink[0]{}%
\providecommand \url  [0]{\begingroup\@sanitize@url \@url }%
\providecommand \@url [1]{\endgroup\@href {#1}{\urlprefix }}%
\providecommand \urlprefix  [0]{URL }%
\providecommand \Eprint [0]{\href }%
\providecommand \doibase [0]{https://doi.org/}%
\providecommand \selectlanguage [0]{\@gobble}%
\providecommand \bibinfo  [0]{\@secondoftwo}%
\providecommand \bibfield  [0]{\@secondoftwo}%
\providecommand \translation [1]{[#1]}%
\providecommand \BibitemOpen [0]{}%
\providecommand \bibitemStop [0]{}%
\providecommand \bibitemNoStop [0]{.\EOS\space}%
\providecommand \EOS [0]{\spacefactor3000\relax}%
\providecommand \BibitemShut  [1]{\csname bibitem#1\endcsname}%
\let\auto@bib@innerbib\@empty
\bibitem [{\citenamefont {Müller}\ \emph {et~al.}(1994)\citenamefont
  {Müller}, \citenamefont {Robaschik}, \citenamefont {Geyer}, \citenamefont
  {Dittes},\ and\ \citenamefont {Hořejši}}]{Muller94}%
  \BibitemOpen
  \bibfield  {author} {\bibinfo {author} {\bibfnamefont {D.}~\bibnamefont
  {Müller}}, \bibinfo {author} {\bibfnamefont {D.}~\bibnamefont {Robaschik}},
  \bibinfo {author} {\bibfnamefont {B.}~\bibnamefont {Geyer}}, \bibinfo
  {author} {\bibfnamefont {F.-M.}\ \bibnamefont {Dittes}},\ and\ \bibinfo
  {author} {\bibfnamefont {J.}~\bibnamefont {Hořejši}},\ }\href
  {https://doi.org/https://doi.org/10.1002/prop.2190420202} {\bibfield
  {journal} {\bibinfo  {journal} {Fortschritte der Physik}\ }\textbf {\bibinfo
  {volume} {42}},\ \bibinfo {pages} {101} (\bibinfo {year} {1994})}\BibitemShut
  {NoStop}%
\bibitem [{\citenamefont {Radyushkin}(1996)}]{Raduyshkin96}%
  \BibitemOpen
  \bibfield  {author} {\bibinfo {author} {\bibfnamefont {A.}~\bibnamefont
  {Radyushkin}},\ }\href
  {https://doi.org/https://doi.org/10.1016/0370-2693(96)00528-X} {\bibfield
  {journal} {\bibinfo  {journal} {Physics Letters B}\ }\textbf {\bibinfo
  {volume} {380}},\ \bibinfo {pages} {417} (\bibinfo {year}
  {1996})}\BibitemShut {NoStop}%
\bibitem [{\citenamefont {Ji}(1997)}]{Ji97}%
  \BibitemOpen
  \bibfield  {author} {\bibinfo {author} {\bibfnamefont {X.}~\bibnamefont
  {Ji}},\ }\href {https://doi.org/10.1103/PhysRevD.55.7114} {\bibfield
  {journal} {\bibinfo  {journal} {Phys. Rev. D}\ }\textbf {\bibinfo {volume}
  {55}},\ \bibinfo {pages} {7114} (\bibinfo {year} {1997})}\BibitemShut
  {NoStop}%
\bibitem [{\citenamefont {Burkardt}(2000)}]{Burkardt00}%
  \BibitemOpen
  \bibfield  {author} {\bibinfo {author} {\bibfnamefont {M.}~\bibnamefont
  {Burkardt}},\ }\href {https://doi.org/10.1103/PhysRevD.62.071503} {\bibfield
  {journal} {\bibinfo  {journal} {Phys. Rev. D}\ }\textbf {\bibinfo {volume}
  {62}},\ \bibinfo {pages} {071503} (\bibinfo {year} {2000})}\BibitemShut
  {NoStop}%
\bibitem [{\citenamefont {Polyakov}(2003)}]{Polyakov03}%
  \BibitemOpen
  \bibfield  {author} {\bibinfo {author} {\bibfnamefont {M.}~\bibnamefont
  {Polyakov}},\ }\href
  {https://doi.org/https://doi.org/10.1016/S0370-2693(03)00036-4} {\bibfield
  {journal} {\bibinfo  {journal} {Physics Letters B}\ }\textbf {\bibinfo
  {volume} {555}},\ \bibinfo {pages} {57} (\bibinfo {year} {2003})}\BibitemShut
  {NoStop}%
\bibitem [{\citenamefont {Belitsky}\ \emph {et~al.}(2002)\citenamefont
  {Belitsky}, \citenamefont {Mueller},\ and\ \citenamefont
  {Kirchner}}]{BMK2002}%
  \BibitemOpen
  \bibfield  {author} {\bibinfo {author} {\bibfnamefont {A.~V.}\ \bibnamefont
  {Belitsky}}, \bibinfo {author} {\bibfnamefont {D.}~\bibnamefont {Mueller}},\
  and\ \bibinfo {author} {\bibfnamefont {A.}~\bibnamefont {Kirchner}},\
  }\bibfield  {title} {\bibinfo {title} {{Theory of deeply virtual Compton
  scattering on the nucleon}},\ }\href
  {https://doi.org/10.1016/S0550-3213(02)00144-X} {\bibfield  {journal}
  {\bibinfo  {journal} {Nucl. Phys. B}\ }\textbf {\bibinfo {volume} {629}},\
  \bibinfo {pages} {323} (\bibinfo {year} {2002})},\ \Eprint
  {https://arxiv.org/abs/hep-ph/0112108} {arXiv:hep-ph/0112108} \BibitemShut
  {NoStop}%
\bibitem [{\citenamefont {Kroll}\ \emph {et~al.}(2013)\citenamefont {Kroll},
  \citenamefont {Moutarde},\ and\ \citenamefont {Sabatie}}]{Kroll:2012sm}%
  \BibitemOpen
  \bibfield  {author} {\bibinfo {author} {\bibfnamefont {P.}~\bibnamefont
  {Kroll}}, \bibinfo {author} {\bibfnamefont {H.}~\bibnamefont {Moutarde}},\
  and\ \bibinfo {author} {\bibfnamefont {F.}~\bibnamefont {Sabatie}},\ }\href
  {https://doi.org/10.1140/epjc/s10052-013-2278-0} {\bibfield  {journal}
  {\bibinfo  {journal} {Eur. Phys. J. C}\ }\textbf {\bibinfo {volume} {73}},\
  \bibinfo {pages} {2278} (\bibinfo {year} {2013})},\ \Eprint
  {https://arxiv.org/abs/1210.6975} {arXiv:1210.6975 [hep-ph]} \BibitemShut
  {NoStop}%
\bibitem [{\citenamefont {Adloff}\ \emph {et~al.}(2001)\citenamefont {Adloff}
  \emph {et~al.}}]{H101}%
  \BibitemOpen
  \bibfield  {author} {\bibinfo {author} {\bibfnamefont {C.}~\bibnamefont
  {Adloff}} \emph {et~al.} (\bibinfo {collaboration} {H1 Collaboration}),\
  }\href {https://doi.org/https://doi.org/10.1016/S0370-2693(01)00939-X}
  {\bibfield  {journal} {\bibinfo  {journal} {Physics Letters B}\ }\textbf
  {\bibinfo {volume} {517}},\ \bibinfo {pages} {47} (\bibinfo {year}
  {2001})}\BibitemShut {NoStop}%
\bibitem [{\citenamefont {Aaron}\ \emph {et~al.}(2008)\citenamefont {Aaron}
  \emph {et~al.}}]{H108}%
  \BibitemOpen
  \bibfield  {author} {\bibinfo {author} {\bibfnamefont {F.}~\bibnamefont
  {Aaron}} \emph {et~al.} (\bibinfo {collaboration} {H1 Collaboration}),\
  }\href {https://doi.org/https://doi.org/10.1016/j.physletb.2007.11.093}
  {\bibfield  {journal} {\bibinfo  {journal} {Physics Letters B}\ }\textbf
  {\bibinfo {volume} {659}},\ \bibinfo {pages} {796} (\bibinfo {year}
  {2008})}\BibitemShut {NoStop}%
\bibitem [{\citenamefont {Chekanov}\ \emph {et~al.}(2003)\citenamefont
  {Chekanov} \emph {et~al.}}]{ZEUS03}%
  \BibitemOpen
  \bibfield  {author} {\bibinfo {author} {\bibfnamefont {S.}~\bibnamefont
  {Chekanov}} \emph {et~al.} (\bibinfo {collaboration} {ZEUS Collaboration}),\
  }\href {https://doi.org/https://doi.org/10.1016/j.physletb.2003.08.048}
  {\bibfield  {journal} {\bibinfo  {journal} {Physics Letters B}\ }\textbf
  {\bibinfo {volume} {573}},\ \bibinfo {pages} {46} (\bibinfo {year}
  {2003})}\BibitemShut {NoStop}%
\bibitem [{\citenamefont {Airapetian}\ \emph {et~al.}(2012)\citenamefont
  {Airapetian} \emph {et~al.}}]{HERMES}%
  \BibitemOpen
  \bibfield  {author} {\bibinfo {author} {\bibfnamefont {A.}~\bibnamefont
  {Airapetian}} \emph {et~al.} (\bibinfo {collaboration} {HERMES
  Collaboration}),\ }\href {https://doi.org/10.1007/JHEP07(2012)032} {\bibfield
   {journal} {\bibinfo  {journal} {Journal of High Energy Physics}\ }\textbf
  {\bibinfo {volume} {2012}},\ \bibinfo {pages} {32} (\bibinfo {year}
  {2012})},\ \Eprint {https://arxiv.org/abs/\textit{See also references
  within}} {\textit{See also references within}} \BibitemShut {NoStop}%
\bibitem [{\citenamefont {Akhunzyanov}\ \emph {et~al.}(2019)\citenamefont
  {Akhunzyanov} \emph {et~al.}}]{COMPASS:2018pup}%
  \BibitemOpen
  \bibfield  {author} {\bibinfo {author} {\bibfnamefont {R.}~\bibnamefont
  {Akhunzyanov}} \emph {et~al.} (\bibinfo {collaboration} {COMPASS}),\ }\href
  {https://doi.org/10.1016/j.physletb.2019.04.038} {\bibfield  {journal}
  {\bibinfo  {journal} {Phys. Lett. B}\ }\textbf {\bibinfo {volume} {793}},\
  \bibinfo {pages} {188} (\bibinfo {year} {2019})},\ \bibinfo {note} {[Erratum:
  Phys.Lett.B 800, 135129 (2020)]},\ \Eprint {https://arxiv.org/abs/1802.02739}
  {arXiv:1802.02739 [hep-ex]} \BibitemShut {NoStop}%
\bibitem [{\citenamefont {Joerg}(2016)}]{COMPASS}%
  \BibitemOpen
  \bibfield  {author} {\bibinfo {author} {\bibfnamefont {P.}~\bibnamefont
  {Joerg}} (\bibinfo {collaboration} {COMPASS}),\ }\href
  {https://doi.org/10.22323/1.265.0235} {\bibfield  {journal} {\bibinfo
  {journal} {PoS}\ }\textbf {\bibinfo {volume} {DIS2016}},\ \bibinfo {pages}
  {235} (\bibinfo {year} {2016})},\ \Eprint {https://arxiv.org/abs/1702.06315}
  {arXiv:1702.06315 [hep-ex]} \BibitemShut {NoStop}%
\bibitem [{\citenamefont {Mu\~noz Camacho}\ \emph {et~al.}(2006)\citenamefont
  {Mu\~noz Camacho} \emph {et~al.}}]{JeffersonLabHallA:2006prd}%
  \BibitemOpen
  \bibfield  {author} {\bibinfo {author} {\bibfnamefont {C.}~\bibnamefont
  {Mu\~noz Camacho}} \emph {et~al.} (\bibinfo {collaboration} {Jefferson Lab
  Hall A, Hall A DVCS}),\ }\href
  {https://doi.org/10.1103/PhysRevLett.97.262002} {\bibfield  {journal}
  {\bibinfo  {journal} {Phys. Rev. Lett.}\ }\textbf {\bibinfo {volume} {97}},\
  \bibinfo {pages} {262002} (\bibinfo {year} {2006})},\ \Eprint
  {https://arxiv.org/abs/nucl-ex/0607029} {arXiv:nucl-ex/0607029} \BibitemShut
  {NoStop}%
\bibitem [{\citenamefont {Defurne}\ \emph {et~al.}(2015)\citenamefont {Defurne}
  \emph {et~al.}}]{HallA15}%
  \BibitemOpen
  \bibfield  {author} {\bibinfo {author} {\bibfnamefont {M.}~\bibnamefont
  {Defurne}} \emph {et~al.} (\bibinfo {collaboration} {Jefferson Lab Hall A
  Collaboration}),\ }\href {https://doi.org/10.1103/PhysRevC.92.055202}
  {\bibfield  {journal} {\bibinfo  {journal} {Phys. Rev. C}\ }\textbf {\bibinfo
  {volume} {92}},\ \bibinfo {pages} {055202} (\bibinfo {year}
  {2015})}\BibitemShut {NoStop}%
\bibitem [{\citenamefont {Defurne}\ \emph {et~al.}(2017)\citenamefont {Defurne}
  \emph {et~al.}}]{HallA17}%
  \BibitemOpen
  \bibfield  {author} {\bibinfo {author} {\bibfnamefont {M.}~\bibnamefont
  {Defurne}} \emph {et~al.} (\bibinfo {collaboration} {Jefferson Lab Hall A
  Collaboration}),\ }\href {https://doi.org/10.1038/s41467-017-01819-3}
  {\bibfield  {journal} {\bibinfo  {journal} {Nature Communications}\ }\textbf
  {\bibinfo {volume} {8}},\ \bibinfo {pages} {1408} (\bibinfo {year}
  {2017})}\BibitemShut {NoStop}%
\bibitem [{\citenamefont {Mazouz}\ \emph {et~al.}(2007)\citenamefont {Mazouz}
  \emph {et~al.}}]{HallA07}%
  \BibitemOpen
  \bibfield  {author} {\bibinfo {author} {\bibfnamefont {M.}~\bibnamefont
  {Mazouz}} \emph {et~al.} (\bibinfo {collaboration} {Jefferson Lab Hall A
  Collaboration}),\ }\href {https://doi.org/10.1103/PhysRevLett.99.242501}
  {\bibfield  {journal} {\bibinfo  {journal} {Phys. Rev. Lett.}\ }\textbf
  {\bibinfo {volume} {99}},\ \bibinfo {pages} {242501} (\bibinfo {year}
  {2007})}\BibitemShut {NoStop}%
\bibitem [{\citenamefont {Benali}\ \emph {et~al.}(2021)\citenamefont {Benali}
  \emph {et~al.}}]{HallA21}%
  \BibitemOpen
  \bibfield  {author} {\bibinfo {author} {\bibfnamefont {M.}~\bibnamefont
  {Benali}} \emph {et~al.} (\bibinfo {collaboration} {Jefferson Lab Hall A
  Collaboration}),\ }\href {https://doi.org/10.1038/s41567-019-0774-3}
  {\bibfield  {journal} {\bibinfo  {journal} {Nature Physics}\ }\textbf
  {\bibinfo {volume} {16}},\ \bibinfo {pages} {191} (\bibinfo {year}
  {2021})}\BibitemShut {NoStop}%
\bibitem [{\citenamefont {Georges}\ \emph {et~al.}(2022)\citenamefont {Georges}
  \emph {et~al.}}]{HallA_prl}%
  \BibitemOpen
  \bibfield  {author} {\bibinfo {author} {\bibfnamefont {F.}~\bibnamefont
  {Georges}} \emph {et~al.} (\bibinfo {collaboration} {Jefferson Lab Hall A
  Collaboration}),\ }\href {https://doi.org/10.1103/PhysRevLett.128.252002}
  {\bibfield  {journal} {\bibinfo  {journal} {Phys. Rev. Lett.}\ }\textbf
  {\bibinfo {volume} {128}},\ \bibinfo {pages} {252002} (\bibinfo {year}
  {2022})}\BibitemShut {NoStop}%
\bibitem [{\citenamefont {Stepanyan}\ \emph {et~al.}(2001)\citenamefont
  {Stepanyan}, \citenamefont {Burkert}, \citenamefont {Elouadrhiri} \emph
  {et~al.}}]{CLAS_DVCS01}%
  \BibitemOpen
  \bibfield  {author} {\bibinfo {author} {\bibfnamefont {S.}~\bibnamefont
  {Stepanyan}}, \bibinfo {author} {\bibfnamefont {V.~D.}\ \bibnamefont
  {Burkert}}, \bibinfo {author} {\bibfnamefont {L.}~\bibnamefont
  {Elouadrhiri}}, \emph {et~al.} (\bibinfo {collaboration} {CLAS
  Collaboration}),\ }\href {https://doi.org/10.1103/PhysRevLett.87.182002}
  {\bibfield  {journal} {\bibinfo  {journal} {Phys. Rev. Lett.}\ }\textbf
  {\bibinfo {volume} {87}},\ \bibinfo {pages} {182002} (\bibinfo {year}
  {2001})}\BibitemShut {NoStop}%
\bibitem [{\citenamefont {Chen}\ \emph {et~al.}(2006)\citenamefont {Chen} \emph
  {et~al.}}]{CLAS_DVCS06}%
  \BibitemOpen
  \bibfield  {author} {\bibinfo {author} {\bibfnamefont {S.}~\bibnamefont
  {Chen}} \emph {et~al.} (\bibinfo {collaboration} {CLAS Collaboration}),\
  }\href {https://doi.org/10.1103/PhysRevLett.97.072002} {\bibfield  {journal}
  {\bibinfo  {journal} {Phys. Rev. Lett.}\ }\textbf {\bibinfo {volume} {97}},\
  \bibinfo {pages} {072002} (\bibinfo {year} {2006})}\BibitemShut {NoStop}%
\bibitem [{\citenamefont {Girod}\ \emph {et~al.}(2008)\citenamefont {Girod}
  \emph {et~al.}}]{CLAS_DVCS08}%
  \BibitemOpen
  \bibfield  {author} {\bibinfo {author} {\bibfnamefont {F.~X.}\ \bibnamefont
  {Girod}} \emph {et~al.} (\bibinfo {collaboration} {CLAS Collaboration}),\
  }\href {https://doi.org/10.1103/PhysRevLett.100.162002} {\bibfield  {journal}
  {\bibinfo  {journal} {Phys. Rev. Lett.}\ }\textbf {\bibinfo {volume} {100}},\
  \bibinfo {pages} {162002} (\bibinfo {year} {2008})}\BibitemShut {NoStop}%
\bibitem [{\citenamefont {Jo}\ \emph {et~al.}(2015)\citenamefont {Jo} \emph
  {et~al.}}]{CLAS_DVCS15}%
  \BibitemOpen
  \bibfield  {author} {\bibinfo {author} {\bibfnamefont {H.~S.}\ \bibnamefont
  {Jo}} \emph {et~al.} (\bibinfo {collaboration} {CLAS Collaboration}),\ }\href
  {https://doi.org/10.1103/PhysRevLett.115.212003} {\bibfield  {journal}
  {\bibinfo  {journal} {Phys. Rev. Lett.}\ }\textbf {\bibinfo {volume} {115}},\
  \bibinfo {pages} {212003} (\bibinfo {year} {2015})}\BibitemShut {NoStop}%
\bibitem [{\citenamefont {Seder}\ \emph {et~al.}(2015)\citenamefont {Seder}
  \emph {et~al.}}]{CLAS_DVCS15_2}%
  \BibitemOpen
  \bibfield  {author} {\bibinfo {author} {\bibfnamefont {E.}~\bibnamefont
  {Seder}} \emph {et~al.} (\bibinfo {collaboration} {CLAS Collaboration}),\
  }\href {https://doi.org/10.1103/PhysRevLett.114.032001} {\bibfield  {journal}
  {\bibinfo  {journal} {Phys. Rev. Lett.}\ }\textbf {\bibinfo {volume} {114}},\
  \bibinfo {pages} {032001} (\bibinfo {year} {2015})}\BibitemShut {NoStop}%
\bibitem [{\citenamefont {Pisano}\ \emph {et~al.}(2015)\citenamefont {Pisano}
  \emph {et~al.}}]{CLAS_DVCS15_3}%
  \BibitemOpen
  \bibfield  {author} {\bibinfo {author} {\bibfnamefont {S.}~\bibnamefont
  {Pisano}} \emph {et~al.} (\bibinfo {collaboration} {CLAS Collaboration}),\
  }\href {https://doi.org/10.1103/PhysRevD.91.052014} {\bibfield  {journal}
  {\bibinfo  {journal} {Phys. Rev. D}\ }\textbf {\bibinfo {volume} {91}},\
  \bibinfo {pages} {052014} (\bibinfo {year} {2015})}\BibitemShut {NoStop}%
\bibitem [{\citenamefont {Hirlinger~Saylor}\ \emph {et~al.}(2018)\citenamefont
  {Hirlinger~Saylor} \emph {et~al.}}]{CLAS_DVCS18}%
  \BibitemOpen
  \bibfield  {author} {\bibinfo {author} {\bibfnamefont {N.}~\bibnamefont
  {Hirlinger~Saylor}} \emph {et~al.} (\bibinfo {collaboration} {CLAS
  Collaboration}),\ }\href {https://doi.org/10.1103/PhysRevC.98.045203}
  {\bibfield  {journal} {\bibinfo  {journal} {Phys. Rev. C}\ }\textbf {\bibinfo
  {volume} {98}},\ \bibinfo {pages} {045203} (\bibinfo {year}
  {2018})}\BibitemShut {NoStop}%
\bibitem [{\citenamefont {Burkert}\ \emph {et~al.}(2020)\citenamefont {Burkert}
  \emph {et~al.}}]{CLAS_NIM}%
  \BibitemOpen
  \bibfield  {author} {\bibinfo {author} {\bibfnamefont {V.}~\bibnamefont
  {Burkert}} \emph {et~al.},\ }\href
  {https://doi.org/https://doi.org/10.1016/j.nima.2020.163419} {\bibfield
  {journal} {\bibinfo  {journal} {Nuclear Instruments and Methods in Physics
  Research Section A: Accelerators, Spectrometers, Detectors and Associated
  Equipment}\ }\textbf {\bibinfo {volume} {959}},\ \bibinfo {pages} {163419}
  (\bibinfo {year} {2020})}\BibitemShut {NoStop}%
\bibitem [{\citenamefont {Masi}\ \emph {et~al.}(2008)\citenamefont {Masi} \emph
  {et~al.}}]{Masi_2008}%
  \BibitemOpen
  \bibfield  {author} {\bibinfo {author} {\bibfnamefont {R.~D.}\ \bibnamefont
  {Masi}} \emph {et~al.} (\bibinfo {collaboration} {CLAS Collaboration}),\
  }\bibfield  {title} {\bibinfo {title} {Measurement of $ep \rightarrow
  ep\pi^0$ beam spin asymmetries above the resonance region},\ }\bibfield
  {journal} {\bibinfo  {journal} {Physical Review C}\ }\textbf {\bibinfo
  {volume} {77}},\ \href {https://doi.org/10.1103/physrevc.77.042201}
  {10.1103/physrevc.77.042201} (\bibinfo {year} {2008})\BibitemShut {NoStop}%
\bibitem [{\citenamefont {Akushevich}\ and\ \citenamefont
  {Ilyichev}(2018)}]{Akusevitch}%
  \BibitemOpen
  \bibfield  {author} {\bibinfo {author} {\bibfnamefont {I.}~\bibnamefont
  {Akushevich}}\ and\ \bibinfo {author} {\bibfnamefont {A.}~\bibnamefont
  {Ilyichev}},\ }\href {https://doi.org/10.1103/PhysRevD.98.013005} {\bibfield
  {journal} {\bibinfo  {journal} {Phys. Rev. D}\ }\textbf {\bibinfo {volume}
  {98}},\ \bibinfo {pages} {013005} (\bibinfo {year} {2018})}\BibitemShut
  {NoStop}%
\bibitem [{\citenamefont {{Kumericki, Kresimir}}\ and\ \citenamefont
  {{M\"uller, Dieter}}(2016)}]{KM15}%
  \BibitemOpen
  \bibfield  {author} {\bibinfo {author} {\bibnamefont {{Kumericki,
  Kresimir}}}\ and\ \bibinfo {author} {\bibnamefont {{M\"uller, Dieter}}},\
  }\href {https://doi.org/10.1051/epjconf/201611201012} {\bibfield  {journal}
  {\bibinfo  {journal} {EPJ Web of Conferences}\ }\textbf {\bibinfo {volume}
  {112}},\ \bibinfo {pages} {01012} (\bibinfo {year} {2016})}\BibitemShut
  {NoStop}%
\bibitem [{\citenamefont {Berthou}\ \emph {et~al.}(2018)\citenamefont {Berthou}
  \emph {et~al.}}]{PARTONS}%
  \BibitemOpen
  \bibfield  {author} {\bibinfo {author} {\bibfnamefont {B.}~\bibnamefont
  {Berthou}} \emph {et~al.},\ }\href
  {https://doi.org/10.1140/epjc/s10052-018-5948-0} {\bibfield  {journal}
  {\bibinfo  {journal} {Eur. Phys. J. C}\ }\textbf {\bibinfo {volume} {78}},\
  \bibinfo {pages} {478} (\bibinfo {year} {2018})},\ \Eprint
  {https://arxiv.org/abs/1512.06174} {arXiv:1512.06174 [hep-ph]} \BibitemShut
  {NoStop}%
\bibitem [{\citenamefont {Moutarde}\ \emph {et~al.}(2019)\citenamefont
  {Moutarde}, \citenamefont {Sznajder},\ and\ \citenamefont
  {Wagner}}]{PARTONSann}%
  \BibitemOpen
  \bibfield  {author} {\bibinfo {author} {\bibfnamefont {H.}~\bibnamefont
  {Moutarde}}, \bibinfo {author} {\bibfnamefont {P.}~\bibnamefont {Sznajder}},\
  and\ \bibinfo {author} {\bibfnamefont {J.}~\bibnamefont {Wagner}},\ }\href
  {https://doi.org/10.1140/epjc/s10052-019-7117-5} {\bibfield  {journal}
  {\bibinfo  {journal} {Eur. Phys. J. C}\ }\textbf {\bibinfo {volume} {79}},\
  \bibinfo {pages} {614} (\bibinfo {year} {2019})},\ \Eprint
  {https://arxiv.org/abs/1905.02089} {arXiv:1905.02089 [hep-ph]} \BibitemShut
  {NoStop}%
\bibitem [{\citenamefont {Giele}\ and\ \citenamefont
  {Keller}(1998)}]{Giele_1998}%
  \BibitemOpen
  \bibfield  {author} {\bibinfo {author} {\bibfnamefont {W.~T.}\ \bibnamefont
  {Giele}}\ and\ \bibinfo {author} {\bibfnamefont {S.}~\bibnamefont {Keller}},\
  }\bibfield  {title} {\bibinfo {title} {Implications of hadron collider
  observables on parton distribution function uncertainties},\ }\bibfield
  {journal} {\bibinfo  {journal} {Physical Review D}\ }\textbf {\bibinfo
  {volume} {58}},\ \href {https://doi.org/10.1103/physrevd.58.094023}
  {10.1103/physrevd.58.094023} (\bibinfo {year} {1998})\BibitemShut {NoStop}%
\bibitem [{\citenamefont {Dutrieux}\ \emph {et~al.}(2021)\citenamefont
  {Dutrieux}, \citenamefont {Bertone}, \citenamefont {Moutarde},\ and\
  \citenamefont {Sznajder}}]{Dutrieux:2021ehx}%
  \BibitemOpen
  \bibfield  {author} {\bibinfo {author} {\bibfnamefont {H.}~\bibnamefont
  {Dutrieux}}, \bibinfo {author} {\bibfnamefont {V.}~\bibnamefont {Bertone}},
  \bibinfo {author} {\bibfnamefont {H.}~\bibnamefont {Moutarde}},\ and\
  \bibinfo {author} {\bibfnamefont {P.}~\bibnamefont {Sznajder}},\ }\bibfield
  {title} {\bibinfo {title} {{Impact of a positron beam at JLab on an unbiased
  determination of DVCS Compton form factors}},\ }\href
  {https://doi.org/10.1140/epja/s10050-021-00560-2} {\bibfield  {journal}
  {\bibinfo  {journal} {Eur. Phys. J. A}\ }\textbf {\bibinfo {volume} {57}},\
  \bibinfo {pages} {250} (\bibinfo {year} {2021})},\ \Eprint
  {https://arxiv.org/abs/2105.09245} {arXiv:2105.09245 [hep-ph]} \BibitemShut
  {NoStop}%
\bibitem [{\citenamefont {Goloskokov}\ and\ \citenamefont
  {Kroll}(2005)}]{GK05}%
  \BibitemOpen
  \bibfield  {author} {\bibinfo {author} {\bibfnamefont {S.~V.}\ \bibnamefont
  {Goloskokov}}\ and\ \bibinfo {author} {\bibfnamefont {P.}~\bibnamefont
  {Kroll}},\ }\href {https://doi.org/10.1140/epjc/s2005-02298-5} {\bibfield
  {journal} {\bibinfo  {journal} {The European Physical Journal C}\ }\textbf
  {\bibinfo {volume} {42}},\ \bibinfo {pages} {281} (\bibinfo {year}
  {2005})}\BibitemShut {NoStop}%
\bibitem [{\citenamefont {Goloskokov}\ and\ \citenamefont
  {Kroll}(2009)}]{GK10}%
  \BibitemOpen
  \bibfield  {author} {\bibinfo {author} {\bibfnamefont {S.~V.}\ \bibnamefont
  {Goloskokov}}\ and\ \bibinfo {author} {\bibfnamefont {P.}~\bibnamefont
  {Kroll}},\ }\bibfield  {journal} {\bibinfo  {journal} {The European Physical
  Journal C}\ }\textbf {\bibinfo {volume} {65}},\ \href
  {https://doi.org/10.1140/epjc/s10052-009-1178-9}
  {10.1140/epjc/s10052-009-1178-9} (\bibinfo {year} {2009})\BibitemShut
  {NoStop}%
\bibitem [{\citenamefont {Guidal}\ \emph {et~al.}(2005)\citenamefont {Guidal},
  \citenamefont {Polyakov}, \citenamefont {Radyushkin},\ and\ \citenamefont
  {Vanderhaeghen}}]{VGG_recent}%
  \BibitemOpen
  \bibfield  {author} {\bibinfo {author} {\bibfnamefont {M.}~\bibnamefont
  {Guidal}}, \bibinfo {author} {\bibfnamefont {M.~V.}\ \bibnamefont
  {Polyakov}}, \bibinfo {author} {\bibfnamefont {A.~V.}\ \bibnamefont
  {Radyushkin}},\ and\ \bibinfo {author} {\bibfnamefont {M.}~\bibnamefont
  {Vanderhaeghen}},\ }\bibfield  {title} {\bibinfo {title} {Nucleon form
  factors from generalized parton distributions},\ }\bibfield  {journal}
  {\bibinfo  {journal} {Physical Review D}\ }\textbf {\bibinfo {volume} {72}},\
  \href {https://doi.org/10.1103/physrevd.72.054013}
  {10.1103/physrevd.72.054013} (\bibinfo {year} {2005})\BibitemShut {NoStop}%
\bibitem [{\citenamefont {Guidal}\ \emph {et~al.}(2013)\citenamefont {Guidal},
  \citenamefont {Moutarde},\ and\ \citenamefont {Vanderhaeghen}}]{VGG_recent2}%
  \BibitemOpen
  \bibfield  {author} {\bibinfo {author} {\bibfnamefont {M.}~\bibnamefont
  {Guidal}}, \bibinfo {author} {\bibfnamefont {H.}~\bibnamefont {Moutarde}},\
  and\ \bibinfo {author} {\bibfnamefont {M.}~\bibnamefont {Vanderhaeghen}},\
  }\bibfield  {title} {\bibinfo {title} {Generalized parton distributions in
  the valence region from deeply virtual compton scattering},\ }\href
  {https://doi.org/10.1088/0034-4885/76/6/066202} {\bibfield  {journal}
  {\bibinfo  {journal} {Reports on Progress in Physics}\ }\textbf {\bibinfo
  {volume} {76}},\ \bibinfo {pages} {066202} (\bibinfo {year}
  {2013})}\BibitemShut {NoStop}%
\end{thebibliography}%

\end{document}